\documentclass [12pt,amsmath,amssymb,preprintnumbers,aps]{revtex4}

\usepackage{graphicx}
\begin{document}
\def\mbf#1{\mbox{\boldmath$#1$}}

\title{A New  Analytical Method for Computing  Solvent Accessible 
Surface Area and its Gradients for Macromolecules}

\author  {Shura Hayryan$^{\rm a}$, Chin-Kun Hu$^{\rm a,*}$, 
Jaroslav Sk\v riv\' anek$^{\rm b}$,  
Edik Hayryan$^{\rm c}$, Imrich Pokorny$^{\rm b}$}

\affiliation{$^{\rm a}$ Institute of Physics, Academia Sinica,
Nankang, Taipei 11529, Taiwan}

\affiliation{$^{\rm b}$Technical University in Ko\v sice, Slovakia}
\affiliation{$^{\rm c}$ Joint Institute for Nuclear Research, Dubna, Russia}

\date{\today}
\maketitle

\noindent{\bf ABSTRACT:} 
In the calculation of thermodynamic properties and three dimensional
structures of macromolecules, such as proteins,  it is important to have
a good algorithm for computing solvent accessible surface area 
of macromolecules. Here we  propose a new analytical method for  
this purpose. In the proposed algorithm,  we consider  the 
 transformation which  maps the spherical circles of intersection of 
atomic surfaces in three-dimensional space onto the circles on  a 
two-dimensional plane and  the problem  of computing solvent 
accessible surface area is transformed  into the problem of computing 
corresponding curve integrals on the plane. 
 This allows to consider only integrals over the circular 
trajectories on the plane. The algorithm is suitable for parallelization. 
Testings on several small proteins have shown a high accuracy 
of the algorithm and a good  performance. 
\vskip 5 mm


\vskip 3cm
\noindent{$^*$}{{\em Correspondence to}: Chin-Kun Hu, E-mail address,
 huck@phys.sinica.edu.tw}

\noindent{}{
Contract/grant sponsor: National Science Council of the 
Republic of China (Taiwan); contract/grant number NSC 92-2112-M-001-063.}
\newpage
\parindent 5mm

{\bf Introduction}

Biological macromolecules, such as DNA, RNA and proteins, are designed by 
the evolutionary process to function in the environment of water 
solutions. Most  important properties of these molecules are related to
 their interactions with the surrounding water molecules. 
Exact determination of the protein-solvent interactions remains a hard 
problem, and different kinds of approximations have been proposed to
solve approximately  the problem. 
One of them is atomic solvation parameters approach propsed by 
Eisenberg and McLachlan \cite{mclachlan} in 1986, in which it has
been assumed that  the solvation energy of atoms or atomic 
groups is proportional to the area of the part of atomic surface 
exposed to the solvent. The total solvation energy of the  whole protein-solvent 
system is then the sum of individual contributions from all atoms \cite{mclachlan}:

\begin{equation}
\Delta G=\sum_i \sigma_iA_i 
\end{equation}
where $A_i$ and  $\sigma_i$ are, respectively, the 
conformation-dependent solvent accessible surface area and
 atomic solvation parameter of 
the atomic group $i$;  and $\sigma_i$ can be
determined  from experiments on model compounds.
The notion of solvent accessible surface was first introduced by Lee and 
Richards \cite{leerich} in 1971. It is defined as the area of the locus of the center of 
probe sphere (solvent molecule) when it rolls on the van der Waals 
surface of the molecule, without penetrating any atom. Actually, it is the area
of the ``van der Waals'' surface of the system in which all radii of the 
atoms $r_i$ are replaced by $r_i+r_{p}$, where $r_p$ is the radius of 
the probe sphere. Another kind of the surface was introduced by 
Richards \cite{richards} in 1977.  It is called molecular surface and represents 
the outside surface of the volume from which the solvent is excluded. 
It consists of two parts: the part of the van der Walls surface to 
which the probe can contact immediately, and the inward looking part 
of the probe surface when it is in contact with  more than one atom 
simultaneously. 
In Figure 1 a system of three atoms $A, B, C$ is schematically 
shown with the probe sphere $P$. The bold dashed line shows the solvent 
accessible surface and the bold continuous line shows the molecular surface. 
In the present work we don't consider the molecular surface. 
 
Calculation of the surface area has always  been a difficult and time consuming work,
especially in the Monte Carlo (MC) or molecular dynamics (MD) simulations
of macromolecules, in which update surface area should be obtained at
every time step.
Thus the search for more efficient methods and advanced algorithms 
is still an important task.    
Generally, two types of algorithms have been developed: analytical 
\cite{connolly83,richmond,connolly85,scheraga1,scheraga2,eisenhaber1, 
braun98} and numerical \cite{shrake,richrich,sila,still,eisenhaber2}. 
For a more complete review on different methods and algorithms the reader 
can consult \cite{braunrev}.  Very recently, Eisenmenger {\it et al.}
\cite{smmp} published a computer package SMMP: a modern package for
simulation of proteins, in which numerical methods are used to
calculate the solvent accessible surface area.  Based on SMMP,
Hayryan {\it et al.} \cite{hhhs01} proposed  a parallel algorithm to calculate 
protein energies and  Lin, Hu and Hansmann \cite{lhh03} used the parallel tempering
method to calculate the specific heat of HP-36 as a function of temperature,
which shows clearly the importance of  including $\Delta G$ of eq. (1)
in the protein energy.  Namely, when $\Delta G$  is included the 
specific heat of HP-36 as a function of temperature shows clealy
a phase transition; otherwise, there is no such a thermal phase transition.

In the present paper we propose a new analytical method for computing
solvent accessible surface area of macromolecules 
and its gradients against the coordinates 
of the atoms. Traditionally, the analytical methods for computing
solvent accessible area and its gradients have been based on the general 
Gauss-Bonnet theorem \cite{carmo}. Here we use a completely different 
approach which maps the three-dimensional (3D)  molecular system onto the plane. 
In this article we describe the method itself and the fortran code ACCAR  
with some benchmark testings on accuracy and performance of the code.  We 
find the results of the testings as quite acceptable when comparing with
results of other method \cite{floudas},  though some 
improvements of the algorithm are possible (see section RESULTS). 
 We can use ACCAR and SMMP package to study the 3D structure and 
thermodynamic properties of proteins in solvent.
Problems for further studies will be discussed in the last section.

\vskip 5 mm
\noindent\rule{6.2in}{0.4mm}

\vskip 5 mm
{\bf Methods}
\vskip 3 mm

{\bf PARAMETRIZATION OF THE SPHERE}

 We describe the protein molecule as a set of  $N$ spherical 
atoms  which can intersect with each other. The spheres are labeled by $N$
integers from  $1$ to $N$.  We define the solvent accessible surface 
of the molecule as the free surface of the system in which the radius 
of  the probe (water molecule) is added to the radii of all atoms. 
  Let $r_i$ be the 
radius of the $i$th sphere, with $(x_i,y_i,z_i)$ being the 
Cartesian coordinates 
of its center. Everywhere in the text it is assumed that $r_i$ 
equals the actual radius of the atom,  plus the radius of the probe sphere.  
Every point $(x,y,z)$ of the sphere satisfies the equation  
\begin{equation}
(x-x_i)^2+(y-y_i)^2+(z-z_i)^2=r_i^{\, 2}. 
\label{isph}
\end{equation} 

Suppose the $i$-th sphere is put tangent to some plane, 
for example, the $z=z_i-r_i$ plane as shown in Figure 2.  
Then every point on the surface of  the sphere can be projected onto the 
plane from its ``top point'' (TP), i.e. the point which is diametrically opposite to the 
tangent point.  This projection creates a continuous one-to-one correspondence
between the points of  sphere and the tangent plane, except the TP.  
For the sake of simplicity, as shown in Figure 2 
we choose the tangent point as the origin of
the tangent plane and $t$  and $s$ axis of the tangent plane to be parallel
to the $x$ and $y$ axis of the sphere, respectively.  The point $(x,~ y,~ z)$
on the sphere is mapped onto the point $(t,~s)$ on the tangent plane. 
Based on Figure 2, we can show easily that
\begin{eqnarray}
t=-2r_i\frac{x-x_i}{z-z_i-r_i}, \nonumber\\
s=-2r_i\frac{y-y_i}{z-z_i-r_i}.
\label{reverse}
\end{eqnarray}
From equations (\ref{isph}) and (\ref{reverse}), we can obtain
following reverse transformation equations: 
\begin{eqnarray}
\label{transfeq}
x=x_i+\frac{4r_i^{\, 2}t}{t^2+s^2+4r_i^{\, 2}},\nonumber\\
y=y_i+\frac{4r_i^{\, 2}s}{t^2+s^2+4r_i^{\, 2}},\\
z=z_i+r_i-\frac{8r_i^{\, 3}}{t^2+s^2+4r_i^{\, 2}}. \nonumber
\end{eqnarray}
As mentioned above, eq. (\ref{transfeq}) holds true for all 
the points on the sphere except the TP  $(x_i,y_i,z_i+r_i)$ .

The intersection of any two spheres is a spherical circle, which is mapped
onto a circle on the tangent plane 
 except the case when the TP is located on 
the intersection circle.  In this case the image of the intersection circle on the 
plane is a straight line.
 All points of the $i$-th sphere which are not located inside  the $j$-th
sphere satisfy eq.  (\ref{isph}) and the  inequality
\begin{equation}
\label{ij}
(x-x_j)^2+(y-y_j)^2+(z-z_j)^2\geq r_j^{\, 2},
\end{equation}
where $r_j$ and $(x_j,y_j,z_j)$ are the radius and the center 
point of the sphere $j$, respectively. 
Combining  (\ref{ij}) and (\ref{transfeq}), we obtain the following 
inequality for the region of points $(t,s)$ which constitute the 
planar image of the points of the sphere $i$ whose distance from the center of 
the $j$-th sphere is larger than or equal to $r_j$ 
\begin{equation}
\label{abcd}
a_j^i(t^2+s^2)+b_j^it+c_j^is+d_j^i\geq 0,
\end{equation}
where

\begin{equation}
\label{ax}
\begin{tabular}{l}
$a_j^i=(x_i-x_j)^2+(y_i-y_j)^2+(z_i+r_i-z_j)^2-r_j^{\, 2}$,\\[3mm]
$b_j^i=8r_i^{\, 2}(x_i-x_j)$,\\[3mm]
$c_j^i=8r_i^{\, 2}(y_i-y_j)$,\\[3mm]
$d_j^i=4r_i^{\, 2}\left[(x_i-x_j)^2+(y_i-y_j)^2+(z_i-r_i-z_j)^2-r_j^{\,
2}\right].$
\end{tabular}
\end{equation}

Inequality (\ref{abcd}) describes a region on the tangent plane  
$\mathbb{R}^2$ which can be one of the following types resulting
from relations of the $i$ and $j$ spheres shown in 
Figure 3.
\begin{itemize} \itemsep -2pt \leftmargini 0pt
\item[a)] {\it An empty set}. The sphere $i$ is fully submerged in the 
sphere $j$.
\item[b)] {\it A single point}. The sphere $i$ is fully inside the sphere
$j$ but contacts it at one single point. 
\item[c)] {\it Interior of a circle} ($a_j^i<0$). Spheres $i$ and $j$ 
overlap, and the TP on the sphere $i$ is inside the sphere $j$.
\item[d)] {\it Half plane} ($a_j^i=0$). The TP belongs to the
intersection circle.
\item[e)] {\it Exterior of a circle} ($a_j^i>0$). 
The TP is outside of the sphere $j$. 
\item[f)] {\it The whole plane} $\mathbb{R}^2$. Spheres $i$ and $j$ don't 
overlap at all. 
\end{itemize}
\vskip 5mm

{\bf CALCULATION OF THE AREA}

Now we show how the area of the part of the spheric surface can be 
calculated by integration along the circle arcs on the tangent plane. 
Let $\Omega_i(x,y,z)$ be the region of the surface on the $i$th sphere 
whose area we want to calculate. Then the area of the region $\Omega_i(x,y,z)$
is given by
\begin{equation}
\label{A1}
A_i=\int\!\!\!\!\!\int\limits_{\!\!\!\!\!\!\Omega_i(x,y,z)}\mid d\mbf{\sigma}\mid,
\end{equation}
where $\mid \!\!d\mbf{\sigma}\!\!\mid$ is the element of the area of the 
surface as shown in Figure 4 and can be expressed in $(t,s)$ coordinates on the 
tangent plane by the equation
\begin{equation}
\label{dsigma}
\mid\!\!d\mbf{\sigma}\!\!\mid=\mid d_t\mbf{r}\times d_s\mbf{r}\mid.
\end{equation}
Here ``$\times$'' means the vector product, $\mbf{r}$ is the radius vector of 
the element $d\mbf{\sigma}$, $d_t\mbf{r}$ and $d_s\mbf{r}$ are its 
differentials with respect to $t$ and $s$, respectively.
Using the transformation formulas  of eq. (\ref{transfeq}), 
we can obtain
\begin{eqnarray}
d_t\mbf{r}=\left(\frac{\partial x}{\partial t}, \frac{\partial y}{\partial t},
\frac{\partial z}{\partial t}\right)dt, \\
\frac{\partial x}{\partial t}=
\frac{4r_i^2(s^2-t^2+4r_i^2)}{(t^2+s^2+4r_i^2)^2}, \\
\frac{\partial y}{\partial t}=-\frac{8r_i^2ts}{(t^2+s^2+4r_i^2)^2}, \\
\frac{\partial z}{\partial t}=\frac{16r_i^3t}{(t^2+s^2+4r_i^2)^2}, \\
\end{eqnarray}
The term $d_s\mbf{r}$ can be calculated in a similar way and eventually we 
obtain 
\begin{equation}
\label{A2}
A_i=\int\!\!\!\!\!\int\limits_{\!\!\!\!\!\!\!\Omega_i(x,y,z)}\mid d\mbf{\sigma}\mid=
16r_i^4\int\!\!\!\int\limits_{\!\!\!\!\Omega_i(t,s)}\frac{dt\,ds}{(t^2+s^2+4r_i^2)^2},
\end{equation}
where the plane region $\Omega_i(t,s)$ represents the image of the spheric region 
$\Omega_i(x,y,z)$ : 
\begin{equation}\label{omega}
\Omega_i(t,s)=\{(t,s)\in \mathbb{R}^2;\quad
a_j^i(t^2+s^2)+b_j^it+c_j^is+d_j^i\geq 0\quad \mbox{for all}\quad
 j\neq i\}.
\end{equation}

 Formula (\ref{omega}) shows that the region $\Omega_i(t,s)$ 
represents 
one of the sets a) - f) described above or the intersection of several 
such sets. In general, the boundary of  $\Omega_i(t,s)$ consists of 
lines and/or circle arcs. 
There is always a possibility to exclude the lines and the circles 
with large radii,
to avoid inaccuracy in computation. This can be done by rotation 
 of the  whole molecule or the relevant part of the molecule by certain 
angle
to bring a more convenient point of the $i$-th sphere as its TP.
Due to this possibility, we can consider only two cases: 
(i) $\Omega(t,s)$ is a 
finite 
region bounded by circle arcs; (ii) $\Omega(t,s)$ represents the whole 
plane without some finite regions bounded by circular arcs. 
Examples for both cases are shown in Figure 5. 


Suppose $\Omega_i(t,s)$ is a bounded region of nonzero measure 
(e.g.  Figure 5a),  
${\cal N}_i$ is the set of  labels
of the spheres which intersect the $i$-th sphere, and
$\Lambda_j^i$ is the number of the arcs which form the boundary of
$\Omega_i$ and ascend from the $j$-th sphere. Now we use 
Green's formula to replace the integration over the plane 
region in eq. (\ref{A2}) by the sum of the integrals over the circular arcs 
along the boundary of the region. The Green's formula says
\begin{equation}
\label{green}
\int\!\!\!\int\limits_{\!\!\!\!\!\!(S)}\left(\frac{\partial Q}{\partial x}-\frac{\partial P}{\partial y}\right)dxdy=\int\limits_{(K)}Pdx+Qdy,
\end{equation}
where functions $P(x,y)$, $Q(x,y)$, and their  partial differentials are continuous functions of $x$ and $y$,
$(K)$ is the positive oriented piecewise smooth 
boundary of the region $(S)$. 
It is easy to see that if we choose 
\begin{eqnarray}
\label{PQ}
P=2r_i^2\frac{-s}{t^2+s^2+4r_i^2}, \nonumber \\ 
Q= 2r_i^2\frac{t}{t^2+s^2+4r_i^2},
\end{eqnarray}
then we obtain from (\ref{A2}) 
\begin{equation}
\label{Ai00}
A_i=\sum\limits_{j\in N_i} \sum\limits_{\lambda=1}^{\Lambda_j^i}
I^i_{j,\lambda},
\end{equation}
where
\begin{equation}\label{Ai1}
I^i_{j,\lambda}=2r_i^2\int\limits_{C_{j,\lambda}^i}
\frac{tds-sdt}{t^2+s^2+4r_i^{\, 2}}.
\end{equation}
Here all arcs $C_{j,\lambda}^i$ together form the boundary of
$\Omega_i$ and are oriented positively with respect to $\Omega_i.$
 In general, each arc $C_{j,\lambda}^i$ is a part of a circle or a line 
but as was mentioned above, we can always avoid the lines.

 Suppose $C_{j,\lambda}^i$ is a circular arc on the plane formed by 
the projection of the intersection circle of  the $i$-th and the $j$-th spheres. 
Then it can be parametrized by the local polar coordinates through the 
following equations: 
\begin{eqnarray}
\label{local}
t=t_0+r_0\cos\varphi,\qquad
s=s_0+r_0\sin\varphi,
\end{eqnarray}
where $t_0=-{b_j^i}/{2a_j^i}$ and $s_0=-{c_j^i}/{2a_j^i}$ 
are the coordinates of the center point of the corresponding circle 
(the planar image of the center of the intersection circle), and 
\begin{equation}
\label{radius}
r_0={\left({\frac{b_j^{i\, 2}+c_j^{i\,
2}-4a_j^id_j^i}{4a_j^{i\, 2}}}\right)}^{1/2}
\end{equation}
is its radius (the planar image of the radius of the intersection circle), 
$\varphi \in [\alpha_{j,\lambda}^i; \beta_{j,\lambda}^i]$, 
$\alpha_{j,\lambda}^i$ and $\beta^{i}_{j,\lambda}$ are the polar angles 
for the vertices of the arc (e.g. $P_1$ and $P_2$ in  Figure 6). 

Therefore, we can write the differentials $ds$ and $dt$ as 
\begin{eqnarray}
\label{dtds}
ds={\left({\frac{b_j^{i\, 2}+c_j^{i\,
2}-4a_j^id_j^i}{4a_j^{i\, 2}}}\right)}^{1/2}\cos \varphi {\,} d\varphi, \nonumber \\
dt=-{\left({\frac{b_j^{i\, 2}+c_j^{i\,
2}-4a_j^id_j^i}{4a_j^{i\, 2}}}\right)}^{1/2}\sin \varphi {\,} d\varphi.
\end{eqnarray}
 By substituting eqs. (\ref{local}) and (\ref{dtds}) in eq. (\ref{Ai1}) and 
replacing the integration along the arc by integration over the 
angle $\varphi$ in the interval
$[\alpha_{j,\lambda}^i; \beta_{j,\lambda}^i ]$, we can obtain
(in order to make the formulas readable, we omit the upper index $i$ in 
the notations of the variables)

\begin{equation}\label{circleint}
I^i_{j,\lambda}=
r_i^2 \left[(\alpha_{j,\lambda}-\beta_{j,\lambda})
\cdot \mbox{sign}(a_j)
+\frac{d_j+4r_i^2a_j}{V_j}
\left(\pi-2\arctan \frac{U_j}{2a_j^2V_j\sin \frac{\beta_{j,\lambda}-
\alpha_{j,\lambda}}2}\right)\right] ,
\end{equation}
where
\begin{eqnarray}
U_j=|a_j|(b_j^{\, 2}+c_j^{\, 2}-2a_jd_j+8r_i^{\, 2}a_j^{\, 2})
\cos\frac{\beta_{j,\lambda}-\alpha_{j,\lambda}}2\nonumber\\
-a_j\left({b_j^{\, 2}+c_j^{\, 2}-4a_jd_j}\right)^{1/2}
\left(b_j\cos \frac{\alpha_{j,\lambda}+\beta_{j,\lambda}}2+c_j\sin
\frac{\alpha_{j,\lambda}+\beta_{j,\lambda}}2\right),
\end{eqnarray}
and

\begin{equation}
V_j=\left[(4r_i^2a_j-d_j)^2+4r_i^2(b_j^2+c_j^2)\right]^{1/2}.
\end{equation}

If the arc $C_{j,\lambda}^i$ is a full circle 
($\beta_j-\alpha_j=2\pi$) then 

\begin{equation}\label{fullcint}
I^i_{j,\lambda}=
2r_i^{\, 2}\pi\cdot \left(-\mbox{sign}(a_j)+
\frac{d_j+4r_i^{\, 2}a_j}{V_j}\right).
\end{equation}
An unbounded area $\Omega_i$ forms the whole plane except some disks. 
In this case integration over the boundary is taken in the 
negative direction (the sum of the integrals $\int\limits_{C_{j,\lambda}^i}
2r_i^2\frac{tds-sdt}{t^2+s^2+4r_i^{\, 2}}$ is negative), and the result is 
added to the area of the whole sphere $4\pi r_i^2$. So, by virtue of 
(\ref{Ai1}) the general formula for surface area is


\begin{equation}\label{Ai2}
A_i=\chi(\Omega_i)+\sum\limits_{j\in N_i}
\sum\limits_{\lambda=1}^{\Lambda_j^i}
\int\limits_{C_{j,\lambda}^i}2r_i^{\, 2}
\frac{tds-sdt}{t^2+s^2+4r_i^{\, 2}} \,\, ,
\end{equation}
where
\begin{equation}
\chi(\Omega_i)=\left\{
\begin{array}{rp{8cm}}
0,&\mbox{$\Omega_i$ is bounded}\\
4\pi r_i^{\, 2}, &\mbox{$\Omega_i$ is the  whole plane except several
disks.}
\end{array}
\right.
\end{equation}

\vskip 10mm

{\bf CALCULATION OF THE GRADIENT}

In equilibrium simulations like Monte Carlo the the behavior of the system is determined
 only by the energy function. But in dynamical studies one needs to calculate also 
the gradient of the energy. The gradient of the solvation energy within the 
solvation parameters model is defined by the gradient of the solvent accessible area. 
In our approach we calculate the gradient of the area with respect 
to the Cartesian coordinates of the atoms in the following way. 
By definition,
the gradient of $A_i$ with respect to the coordinate $(x_l,y_l,z_l)$
 is
\begin{equation}
\label{denot}
\frac{\partial A_i}{\partial(x_l,y_l,z_l)}\stackrel{def}{=}
\left( \frac{\partial A_i}{\partial x_l};\frac{\partial A_i}
{\partial y_l};\frac{\partial A_i}{\partial z_l}\right), 
\end{equation} 
which can be calculated by the chain rule:
\begin{equation}
\frac{\partial A_i}{\partial (x_l,y_l,z_l)}=
\frac{\partial A_i}{\partial (a_j^i,b_j^i,c_j^i,d_j^i)}\cdot
\frac{\partial (a_j^i,b_j^i,c_j^i,d_j^i)}{\partial (x_l,y_l,z_l)}.
\end{equation} 
If $l\neq j$ and $l \neq i$,  then 
\begin{equation}
\frac{\partial (a_j^i,b_j^i,c_j^i,d_j^i)}{\partial (x_l,y_l,z_l)}=0.
\end{equation} 
If $l=j$,  then using eqs. (\ref{ax}) we  can obtain  
\begin{equation}
\frac{\partial (a_j^i,b_j^i,c_j^i,d_j^i)}{\partial (x_j,y_j,z_j)}=
\left(
\begin{array}{ccc}
2(x_j-x_i)&2(y_j-y_i)&2(z_j-z_i-r_i)\\
-8r_i^{\, 2}&0&0\\
0&-8r_i^{\, 2}&0\\
8r_i^{\, 2}(x_j-x_i)&8r_i^{\, 2}(y_j-y_i)&8r_i^{\,
2}(z_j-z_i+r_i)
\end{array}
\right)
\end{equation}
If  the $j$-th circle has  $\Lambda_j^i$ arcs on the border of $\Omega_i$ 
 (see eq. (\ref{Ai1})), then we have
\begin{equation}\label{DA}
\frac{\partial A_i}{\partial (a_j^i,b_j^i,c_j^i,d_j^i)}=
\sum\limits_{\lambda=1}^{\Lambda_j^i}
\frac{\partial (I_{j,\lambda}^i+I_{k_{\lambda},\mu_{\lambda}}^i+
I_{l_{\lambda},\nu_{\lambda}}^i)}{\partial
(a_j^i,b_j^i,c_j^i,d_j^i)}\,\, .
\end{equation}
Here the integration path of $I_{k_{\lambda},\mu_{\lambda}}$ is the part of
the $k_{\lambda}$-th circle on the border of $\Omega_i$, which 
takes up $C_{j,\lambda}^i$ in the initial point,
and the integration path of  $I_{l_{\lambda},\nu_{\lambda}}^i$ is the part of
the $l_{\lambda}$-th circle, which takes up
$C_{j,\lambda}^i$ in the endpoint. The term $I_{j,\lambda}^i$ in the 
numerator of the right side of eq. (\ref{DA}) depends on the variables
 $(a_j^i,b_j^i,c_j^i,d_j^i)$ explicitly, while the dependence of the two 
other therms on the same variables is complex. Therefore, the 
calculation techniques for these two types of therms are different. 
The details on calculation of the terms
$\frac{\partial I_{k_{\lambda},\mu_{\lambda}}^i}{\partial
(a_j^i,b_j^i,c_j^i,d_j^i)}$ and 
$\frac{\partial I_{l_{\lambda},\nu_{\lambda}}^i}{\partial
(a_j^i,b_j^i,c_j^i,d_j^i)}$ are given at the end of the next section. 
The derivatives of $I_{j,\lambda}^i$ are calculated 
by using the equations (\ref{fullcint}) and 
(\ref{circleint}), as follows. 
When the $j$-th circle on the
border of $\Omega_i$ contains no vertices (i.~e. $\Lambda_j^i=1$ and $C_{j,1}^i$ is a full circle) then 
the derivatives of $I_{j,1}^i$ with respect to the coordinates of an arbitrary 
atom $l$ are 
\begin{equation}
\frac{\partial I_{j,1}^i}{\partial (a_j^i,b_j^i,c_j^i,d_j^i)}=
\frac{8r_i^{\, 4}\pi}{\left((4r_i^{\, 2}a_j^i-d_j^i)^2+
4r_i^{\, 2}(b_j^{i\, 2}+c_j^{i\, 2})\right)^{\frac32}}\cdot
\left(4r_i^{\, 2}(b_j^{i\, 2}+c_j^{i\, 2}-2a_j^id_j^i)
\right.
\end{equation}
$$\left. +2d_j^{i\, 2},-b_j^i(d_j^i+4r_i^{\, 2}a_j^i),-c_j^i(d_j^i+4r_i^{\, 2}a_j^i),
b_j^{i\, 2}+c_j^{i\, 2}-2a_j^id_j^i+8r_i^{\, 2}a_j^{i\,
2}\right).$$
Equation (\ref{fullcint}) was used in this derivation.
For the case of the partial arcs a similar formulas can be derived by using eq. (\ref{circleint}).

Eventually, for $l=i$, i.e. for the gradient of the area $A_i$ with 
respect to the ``own'' coordinates $(x_i,y_i,z_i)$, we have 

\begin{equation}\label{grad_i}
\frac{\partial A_i}{\partial (x_i,y_i,z_i)}=-\sum_{j\in N_i}
\frac{\partial A_i}{\partial (x_j,y_j,z_j)}.
\end{equation}

\vskip 5 mm
\noindent\rule{6.2in}{0.4mm}

\vskip 5mm
{\bf Results}
\vskip 5mm
{\bf DESCRIPTION OF THE ALGORITHM FOR SURFACE CALCULATIONS}

The accessible surface area $A$ of the whole molecule can be calculated 
from the summation of  $A_i$ over $N$ spheres, where $1 \le i \le N$.  
Each particular $A_i$ is calculated in several steps. First the algorithm reads
 the coordinates and radii of the atoms. In the preparatory step 
for each sphere $i$ the computer algorithm generates a list of spheres, 
which contain the labels of the spheres which reach the $i$-th sphere. 
We will call it a list of  ``local spheres''. If some sphere cover in full the
$i$-th sphere then the calculation of $A_i$ is stopped with the result $A_i=0.$


The further cycle compares each pair of relevant regions of eq. (\ref{abcd}).
The region which is a superset of another is removed from the list. If the 
intersection of two regions is of measure zero then we put $A_i=0$ again. In
this cycle the vertices are calculated.
Let the two vertices
$\left[t_{j,k,1};s_{j,k,1}\right]$ and $\left[ t_{j,k,2},s_{j,k,2}\right]$
come from the intersection of borders of the regions $j$ and $k.$
If these borders are circles, then for $\nu=1$ and $\nu=2$ we have

\begin{eqnarray}
t_{j,k,\nu}=\frac{-2(b_j'-b_k')(d_j'-d_k')-(b_k'c_j'-b_j'c_k')
(c_j'-c_k')\pm (c_j'-c_k')\sqrt{D_{j,k}}}{2\left((b_j'-b_k')^2+
(c_j'-c_k')^2\right)},\nonumber\\
s_{j,k,\nu}=\frac{-2(c_j'-c_k')(d_j'-d_k')-(c_k'b_j'-c_j'b_k')
(b_j'-b_k')\mp (b_j'-b_k')\sqrt{D_{j,k}}}{2\left((b_j'-b_k')^2+
(c_j'-c_k')^2\right)},
\end{eqnarray}
where 
 $b_j'=b_j/a_j,~c_j'=c_j/a_j,~d_j'=d_j/a_j, 
~b_k'=b_k/a_k, ~c_k'=c_k/a_k, ~d_k'=d_k/a_k$,  and
\begin{equation}
D_{j,k}=4(b_k'd_j'-b_j'd_k')(b_j'-b_k')
+4(c_k'd_j'-c_j'd_k')(c_j'-c_k')-4(d_j'-d_k')^2+
(c_k'b_j'-c_j'b_k')^2.
\end{equation}

If $\left[t_{j,k};s_{j,k}\right]$ is an initial point or an endpoint
of an arc on the $j$-th circle then

\begin{equation}
\tan \gamma=\frac{2a_j^i\cdot s_{j,k}+c_j^i}{2a_j^i\cdot
t_{j,k}+b_j^i}, 
\end{equation}
where $\gamma=\alpha_j^i$ or $\gamma=\beta_j^i$, respectively.
Therefore, we build up two lists: 
\begin{itemize} \itemsep -2pt \leftmargini 0pt
\item[a)] the list of circles with no vertex;

\item[b)] the list of partial boundary arcs.
\end{itemize}
The whole set of the intersection points on certain circle, ordered according to 
the circle's orientation, defines the set of all possible arcs on it (see Figure 5). 
Only the subset of these arcs corresponds to the part of boundary 
of $\Omega_i$. Only the arcs, which are inside of all other positively 
oriented circles and outside of all other negatively oriented circles with respect to
$\Omega_i$, are included in the list of arcs.   
The procedure is completed by integration over the relevant circles and arcs  
by eqs. (\ref{circleint}) and (\ref{fullcint}), 
 and are summarized by eq. (\ref{Ai2}). 

Now we evaluate a useful implicit equation for  
 computing  the terms
$\frac{\partial I_{k_{\lambda},\mu_{\lambda}}^i}{\partial
(a_j^i,b_j^i,c_j^i,d_j^i)}$ and 
$\frac{\partial I_{l_{\lambda},\nu_{\lambda}}^i}{\partial
(a_j^i,b_j^i,c_j^i,d_j^i)}$
 in  the right side of eq. (\ref{DA}).
Consider the circle 
\begin{equation}
\label{circ1}
a_j(t^2+s^2)+b_j t+c_j s+d_j=0
\end{equation}
in the parametric form
\begin{eqnarray}
\label{para}
t=\frac{-b_j}{2a_j}+\frac{(b_j^2+c_j^2-4a_jd_j)^{1/2}}{2|a_j|}\cdot\cos(\gamma),\nonumber \\
s=\frac{-c_j}{2a_j}+\frac{(b_j^2+c_j^2-4a_jd_j)^{1/2}}{2|a_j|}\cdot\sin(\gamma).
\end{eqnarray}
By substituting these $t$ and $s$ into the equation of the $k$-th circle 
\begin{equation}
\label{circ2}
a_k(t^2+s^2)+b_k t+c_k s+d_k=0,
\end{equation}
and carrying out some short calculations, one obtains 
\begin{eqnarray}\label{Dalpha}
(a_jb_k-a_kb_j)^2+(a_jc_k-a_kc_j)^2+(b_j^{\, 2}+c_j^{\,
2}-4a_jd_j)\cdot a_k^{\, 2}\nonumber\\
-(b_k^{\, 2}+c_k^{\, 2}-4a_kd_k)\cdot a_j^{\, 2}+
2\sqrt{b_j^{\, 2}+c_j^{\, 2}-4a_jd_j}\cdot a_k\nonumber\\
\times \mbox{sign } a_j\cdot \left((a_jb_k-a_kb_j)\cos \gamma+(a_jc_k-a_kc_j)
\sin \gamma \right)=0.
\end{eqnarray}
 This equation allows to avoid  multiple direct calculations of the 
coordinates 
of the intersection points of two arcs during the derivation of the 
differentials on the right side of eq. (\ref{DA}). We use this equation 
in our computer algorithm for gradients.

\vskip 10mm

{\bf TESTINGS FOR ACCURACY AND PERFORMANCE}

 The first test for the accuracy of the method was by calculating the 
accessible area of artificial systems consisting of up to $5$ spheres 
for which 
the calculations ``by hand'' are possible. Many configurations with 
different level of overlapping were used. For all configurations of 
these systems we obtained absolutely exact results.  
 Then we have checked the accuracy of our method by calculating the solvation 
energy ($E_{HYD}$) of small peptide Met-Enkephalin in several  
conformations with different sets of solvation parameters. We have compared 
the obtained results with the data from Tables VIII and IX in \cite{floudas}. 
The comparative data are shown in Table~\ref{table1}.

\begin{table}[ht]
\caption{}\label{table1}

\begin{tabular}{c||r|r}
Solvation & \multicolumn{2}{|c}{$E_{HYD}$ (kcal/mol)}\\ \cline{2-3}          
 model  &       This method &    Reference work \\ \hline
 \verb+OONS+  &           $-22.23$\verb+    +& $-22.23$\verb+    +\\
 \verb+ JRF+  &           $-287.17$\verb+    +& $-288.82$\verb+    +\\
 \verb+ WE1+  &           $-21.80$\verb+    +& $-20.56$\verb+    +\\
 \verb+ WE2+  &           $-24.59$\verb+    +& $-23.49$\verb+    +\\
 \verb+SCKS+  &           $9.242$\verb+    +& $11.00$\verb+    +\\
\end{tabular}
\end{table} 
One can see a very good agreement in the first line of Table I. 
The slight differences in remaining lines result mainly from the 
fact that 
in our calculations we use the ECEPP/3 forcefield which is most 
consistent with OONS set. Another not significant source of 
differences may be the numerical nature of the algorithm used by the 
authors of  Ref.  \cite{floudas}.

 Another test for the performance  of  our algorithm with
computer package  ACCAR was done by calculating the solvent 
accessible area  for several well known peptides. 
 Table II shows the computing times for each peptide in milliseconds.  
\begin{table}[ht]
\caption{}\label{table2}

\begin{tabular}{r||r|r}        
 Name/PDB code  &       Number of atoms &    Time (msec) \\ \hline
 \verb+L-protein+  &           $476$\verb+    +& $101$\verb+    +\\
 \verb+ crambin+  &           $327$\verb+    +& $65$\verb+    +\\
 \verb+ kalata-b1+  &           $198$\verb+    +& $36$\verb+    +\\
 \verb+ H36+  &           $295$\verb+    +& $57$\verb+    +\\
 \verb+4pti+  &           $454$\verb+    +& $95$\verb+    +\\
\end{tabular}
\end{table}

Like any other analytical method, ACCAR is slower than numerical 
algorithms for low accuracy calculations.
For example, it is $3$ times slower than ENYSOL in SMMP \cite{smmp}, 
which utilizes double cubic lattice method with $122$ surface dots.  
However, for very high accuracy calculations our analytic  method can be faster
than numerical methods.
 We failed to compare directly the performance of ACCAR with other 
analytical algorithms because we don't have any at hand. Comparison 
with the published data such as  those reported in \cite{vB} and  \cite{eis}, etc,
 seems not  reliable because different computational platforms are used. For this 
reason we bring our benchmark results without comparing to the others. 
 All calculations are done on PC Pentium II 333MHZ.
\vskip 5 mm
\noindent\rule{6.2in}{0.4mm}

{\bf Concluding remarks and further work}

In this paper, we present
a new analytical method  for computing the solvent 
accessible surface area of biomolecular surface and its gradients. 
The method is tested 
for accuracy and has shown a good agreement with the published data as 
well as with the direct numerical calculations for artificial small systems. 
The ACCAR routine is written and  can be incorporated into the protein 
simulation package SMMP \cite{smmp}.  Calculation of the accessible area for each 
atom is done independently, which provides a natural way for parallel 
realization of the algorithm. This will be a work for the near future. 
Also the efficiency of the gradient calculation part of ACCAR must be 
improved.  
This is the first version of the ACCAR routine 
which was written mainly for checking the accuracy of the method and 
hence, not complete attention was paid to its efficiency. We believe 
that obtained formulas are simple, and the interested reader can easily 
use them to create his own computer code.   

The method presented in this paper can be applied  to calculate
$\Delta G$  of eq. (1),  which can be used  in the 
multicanonical simulation \cite{hhhs01} or
parallel tempering simulation of proteins \cite{lhh03}. 
Our research in this direction is in progress.

\newpage
\centerline{Figure Caption}
\noindent{}FIGURE 1.  ~ Definition of two kinds of surfaces for  a 
system consisting of three atoms $A$, $B$, and $C$. The bold continuous line is the molecular 
surface. The bold dashed line shows the solvent accessible surface. 
$P$  is the probe sphere  and represents the water molecule.  
\vskip 5 mm
\noindent{}FIGURE 2. ~The projection of the point $(x,y,z)$ on the sphere 
onto the point $(t,s)$ on the tangent plane .
\vskip 5mm
\noindent{}FIGURE 3. ~Different relative locations of two spheres 
$i$ and $j$,  which create six different patterns on the tangent plane described in 
the text. 
\vskip 5mm
\noindent{}FIGURE 4. ~The element $d{\mbf \sigma}$ of the spherical surface.
\vskip 5mm
\noindent{}FIGURE 5. ~a) The region  $\Omega_i(t,s)$  on the tangent plane
is bounded  by three circle arcs with vertices $P_1, P_2, P_3$. 
Integration goes in the counterclockwise 
direction (positive with respect to $\Omega$), and the result is a positive quantity. 
b) $\Omega_i(t,s)$ 
represents an unbounded plane area , namely the whole plane without two 
bounded regions. Integration goes over the circle arcs of the bounded regions
in clockwise direction (negative with respect to $\Omega$), and the result 
is negative. 
\vskip 5mm
\noindent{}FIGURE 6. ~ A circular arc corresponding to the angular 
interval $[\alpha, \beta]$.

\newpage
\begin{figure}[ht]
\centering
\includegraphics{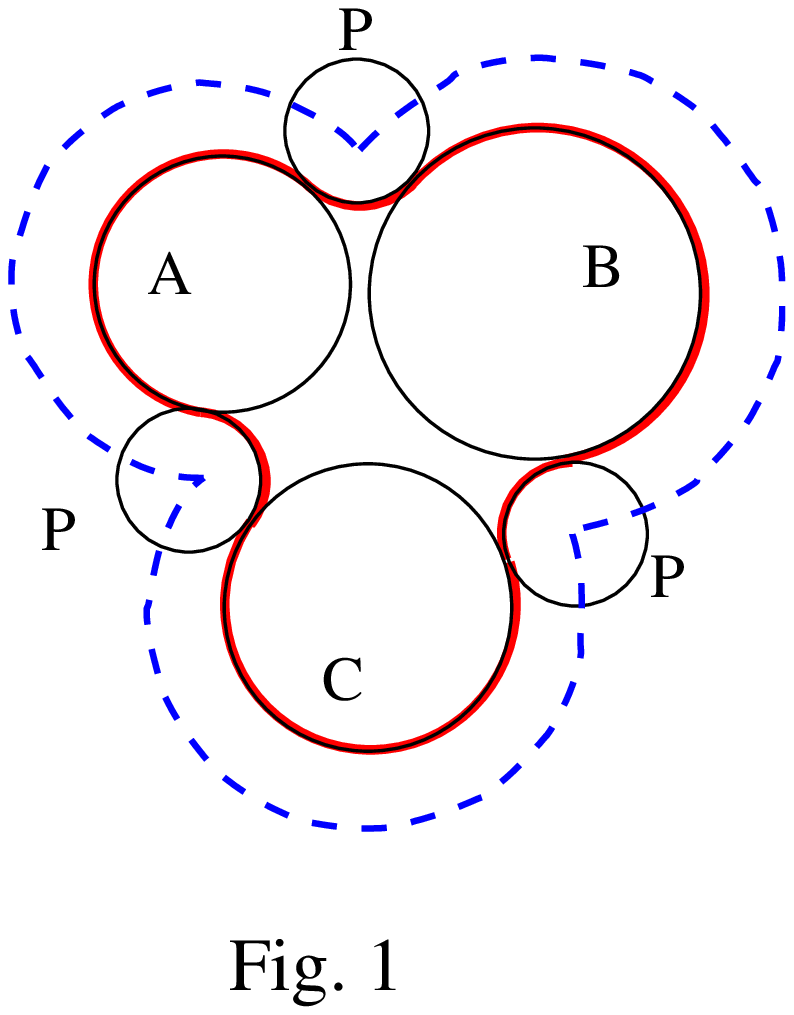}
\label{area1}
\end{figure}

\newpage
\begin{figure}[ht]
\centering
\includegraphics{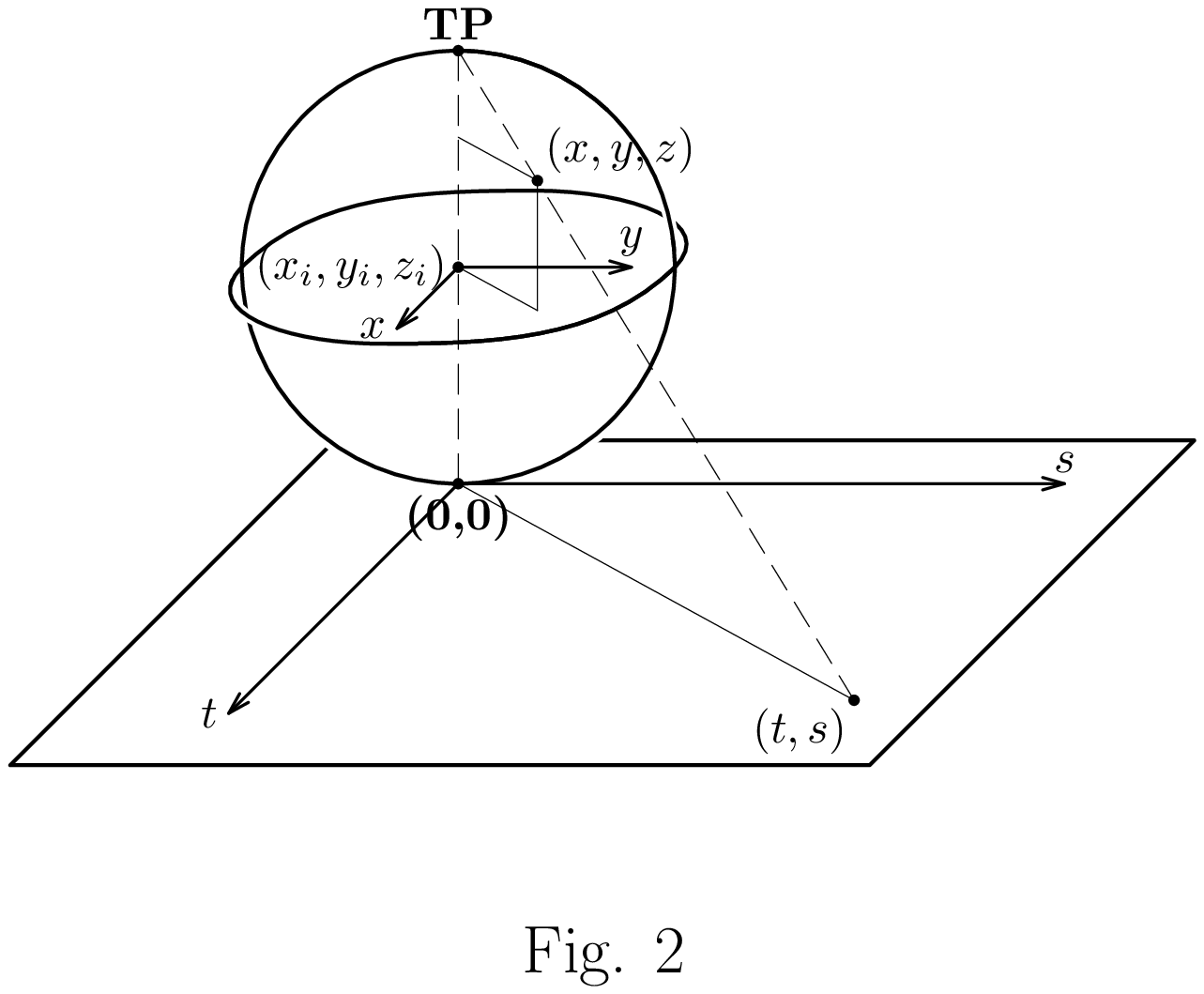}
\label{area2}
\end{figure}

\newpage
\begin{figure}[ht]
\centering
\includegraphics[height=25cm]{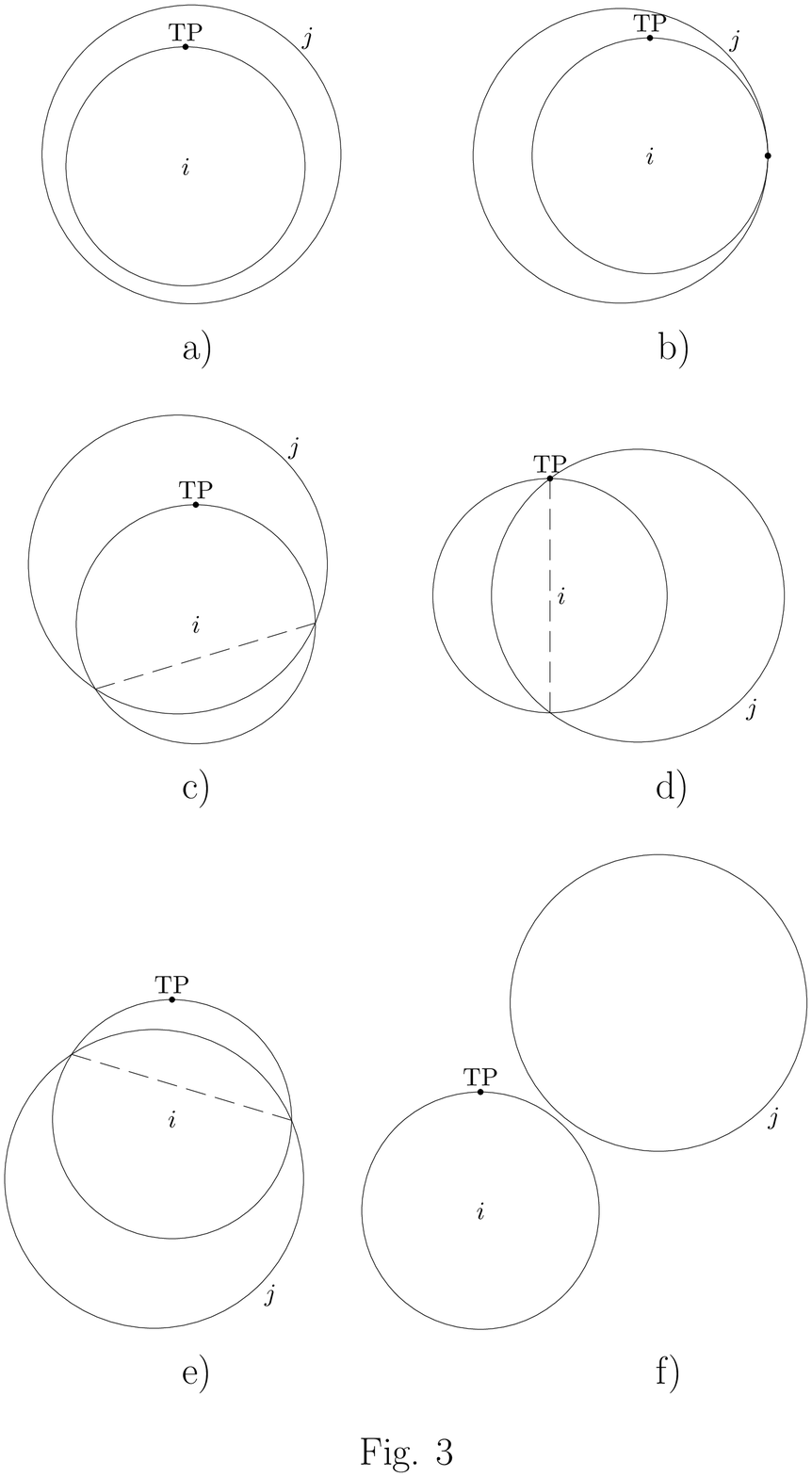}
\label{area3}
\end{figure}

\newpage
\begin{figure}[ht]
\centering
\includegraphics{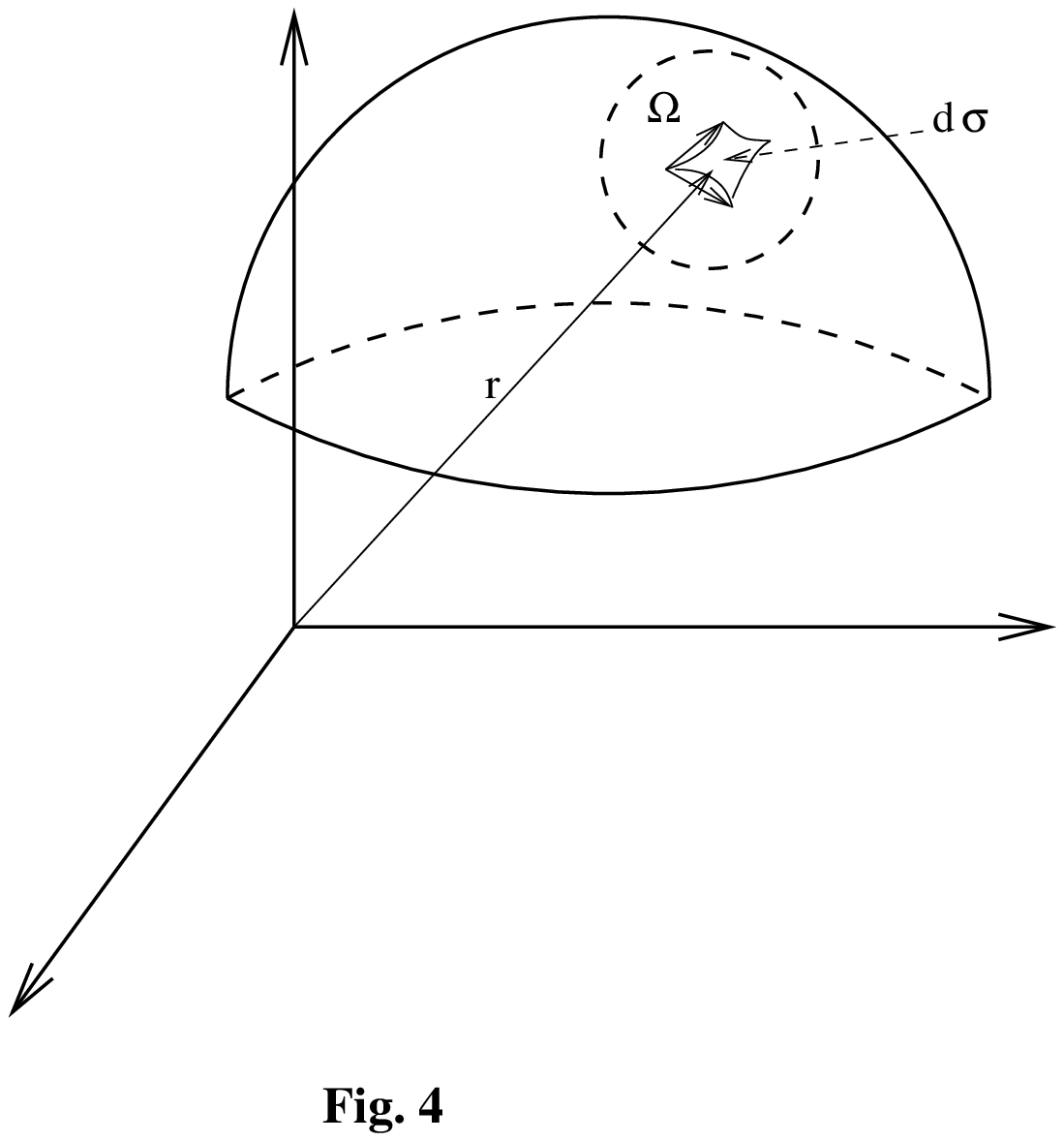}
\label{area4}
\end{figure}

\newpage
\begin{figure}[ht]
\centering
\includegraphics[height=25cm]{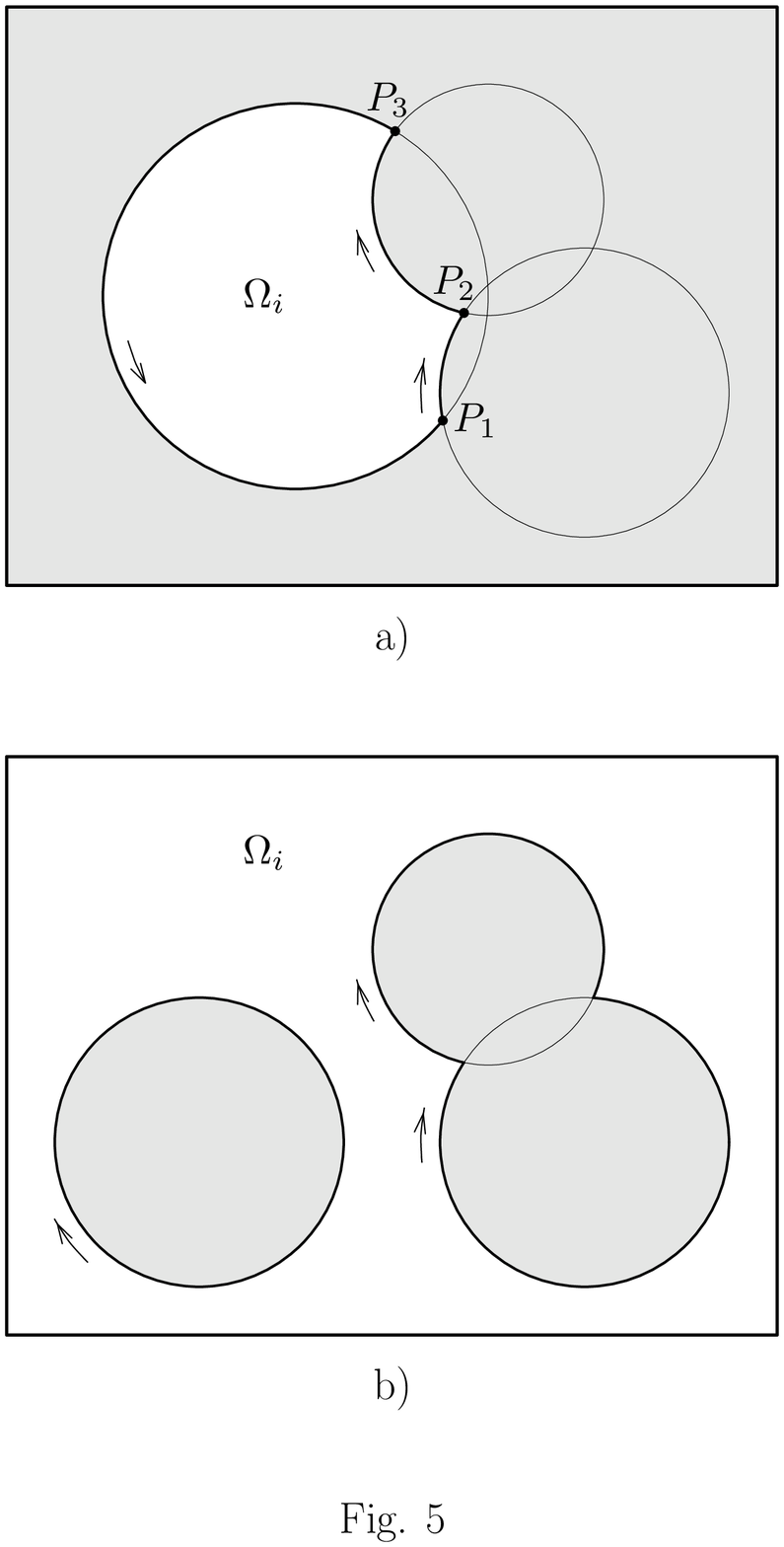}
\label{area5}
\end{figure}

\newpage
\begin{figure}[ht]
\centering
\includegraphics{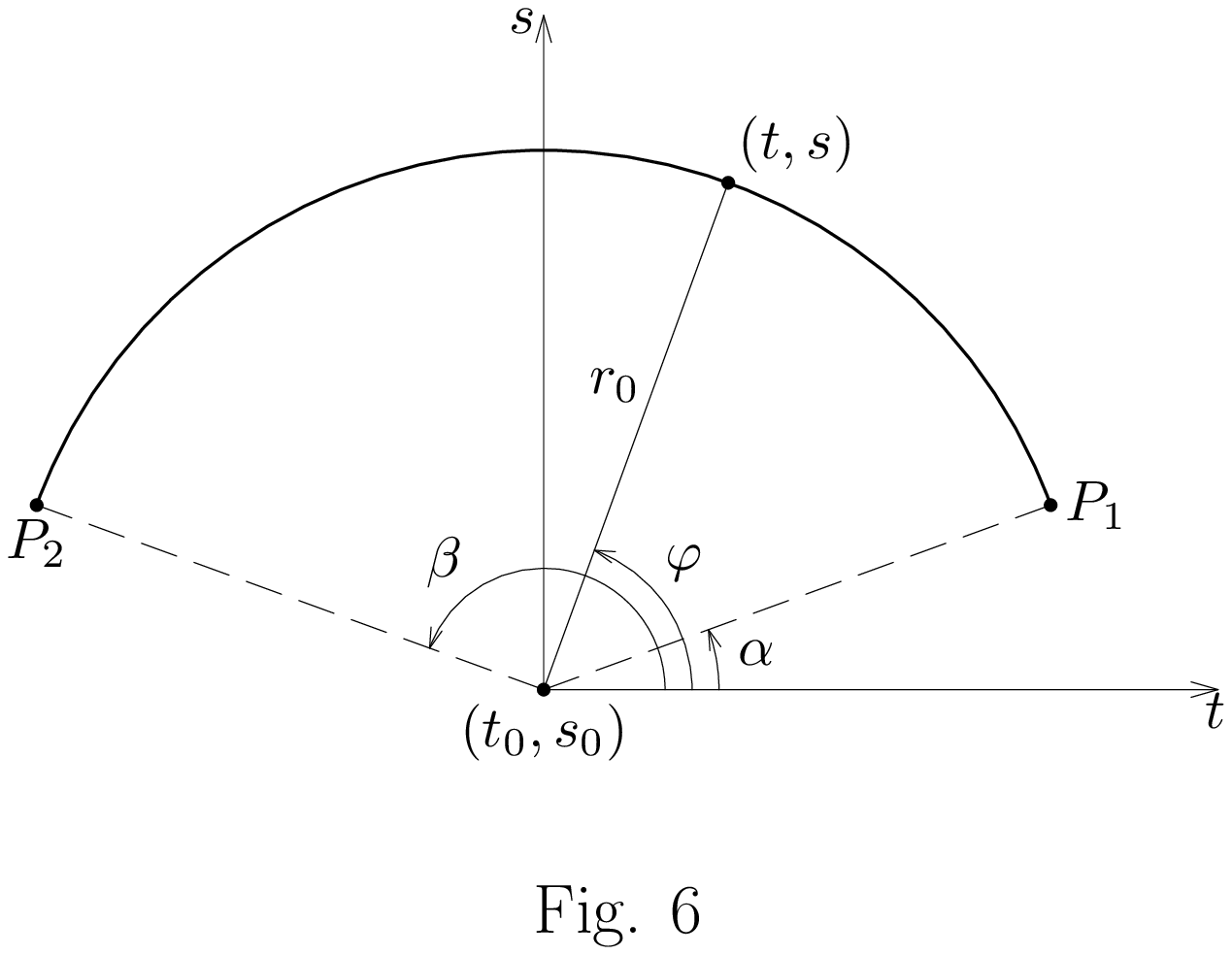}
\label{area6}
\end{figure}

\newpage
\begin{thebibliography}{9}

\bibitem{mclachlan} Eisenberg, D.; McLachlan, A. D. Nature 1986, 316, 199.
\bibitem{leerich} Lee, B.; Richards, F. M. J Mol Biol 1971, 55, 379. 
\bibitem{richards} Richards, F. M. Ann Rev Biophys Bioeng 1977, 6,151.  
\bibitem{connolly83} Connolly, M. L. J Appl Cryst 1983, 16, 548.
\bibitem{richmond} Richmond, T.J. J Mol Biol 1984, 178, 63.
\bibitem{connolly85} Connolly, M. L. J Appl Cryst 1985, 18, 499.
\bibitem{scheraga1} Gibson, K. D.; Scheraga, H. A. Mol Physics 1988, 64, 641.
\bibitem{scheraga2} Perrot, C.; Cheng, B.; Gibson, K. D.; Vila, J.; 
Palmer, K. A.; Nayeem, A.; Maigret, B.; Scheraga, H. A. J Comp Chem 1992, 13, 1.
\bibitem{eisenhaber1} Eisenhaber, F.; Argos, P. J Comp Chem 1993, 14, 1272. 
\bibitem{braun98} Fraczkiewicz, R.; Braun, W. J Comp Chem 1998, 19, 319. 

\bibitem{shrake} Shrake, A.; Rupley, J. A. J Mol Biol 1973, 79, 351.
\bibitem{richrich} Richmond, T. J.; Richards, F. M. J Mol Biol 1978, 129, 527.
\bibitem{sila} Sila, E.; Tunom, I.; Pascual-Ahuir, J. L. J Comp Chem 1991, 12, 1077.
\bibitem{still} Still, W. C.; Tempczyk, A.; Hawley, R.C.; 
Hendricson, T. J Am Chem Soc 1990, 112, 6127.
\bibitem{eisenhaber2} Eisenhaber, F.; Lijnzaad, Ph.; Argos, P.; Sander, C.; 
Scharf, M. J Comp Chem 1995, 16, 273.

\bibitem{braunrev} Braun, W. In {\em Computer Simulation of 
Biomolecular Systems}, v.3, van Gunsteren, W.; Weener, P; Wilkinson, 
T. Eds. ESCOM, Laiden, 1996.

\bibitem{smmp} Eisenmenger, F.; Hansmann, U. H. E.; Hayryan, Sh.; Hu, C.-K. Comp Phys Comm 2001, 138, 193.

\bibitem{hhhs01} Hayryan, S.; Hu, C.-K.; Hu, S.-Y.; Shang, R.-J, J Comp Chem 2001, 22, 1287.

\bibitem{lhh03} Lin, C.-Y.; Hu, C.-K.; Hansmann, U. H. E. Proteins: Structure, Function and Genetics 2003,
  52, 436.


\bibitem{carmo} Carmo, M. P. do In {\em Differential Geometry of 
Curves and Surfaces}, Englewood Cliffs, New Jersey: Prentice-Hall, 1976.
\bibitem{floudas} Klepeis, J. L.; Floudas, C. A. J Comp Chem 1999, 20, 636.
\bibitem{vB} von Freyberg, B.; Braun, W. J Comp Chem 1993, 14, 510.
\bibitem{eis} Eisenhaber, F.; Lijnzaad, P.; Argos, P.; Sander, Ch.; Schaff, M. J Comp Chem 1995, 16, 273.

\end{thebibliography}
\end{document}